\documentclass[a4paper, times, 10pt,twocolumn]{article}
\usepackage[top=4.9cm,bottom=3.7cm,left=1.5cm,right=1.5cm]{geometry}
\usepackage{ICMLC}
\usepackage{times}
\usepackage{graphicx}
\usepackage{indentfirst}
\usepackage{latexsym}
\usepackage{amsmath}
\usepackage{amssymb}
\usepackage{graphicx} 

\usepackage{subfigure}
\usepackage{hyperref}
\usepackage{algorithm}
\usepackage{algpseudocode}

\usepackage[margin=8pt,font=footnotesize,labelfont=bf,labelsep=period
]{caption}

\pdfpagewidth=\paperwidth
\pdfpageheight=\paperheight
\begin{document}

\title{LEARNING TO COMMUNICATE IN MULTI-AGENT REINFORCEMENT LEARNING FOR AUTONOMOUS CYBER DEFENCE}  

\author{\bf{\normalsize{FAIZAN CONTRACTOR${^1}$, LI LI${^2}$}, RANWA AL MALLAH${^1}$}\\ 
\\
\normalsize{$^1$Electrical and Computer Engineering Department, Royal Military College of Canada, Kingston, Ontario, Canada}\\
\normalsize{$^2$Defence Research and Development Canada, Ottawa, Ontario, Canada} \\
\normalsize{E-MAIL: faizan.contractor@gmail.com, li.li2@ecn.forces.gc.ca, ranwa.al-mallah@polymtl.ca}\\
\\}

\maketitle \thispagestyle{empty}

\begin{abstract}
   {Popular methods in cooperative Multi-Agent Reinforcement Learning with partially observable environments typically allow agents to act independently during execution, which may limit the coordinated effect of the trained policies. However, by sharing information such as known or suspected ongoing threats, effective communication can lead to improved decision-making in the cyber battle space. We propose a game design where defender agents learn to communicate and defend against imminent cyber threats by playing training games in the Cyber Operations Research Gym, using the Differentiable Inter Agent Learning algorithm adapted to the cyber operational environment. The tactical policies learned by these autonomous agents are akin to those of human experts during incident responses to avert cyber threats. In addition, the agents simultaneously learn minimal cost communication messages while learning their defence tactical policies.}
\end{abstract}

\Section{Introduction}

In recent years, the rapid advancement of autonomous agent technology across various application domains \cite{Peter} has led to experimentation with Autonomous Cyber Defence (ACD) agents, aiming to achieve machine-speed scalability in cyber defence operations. ACD agents apply Deep Reinforcement Learning (DRL) techniques, similar to agents being developed for other applications \cite{Wiebe} \cite{Garrett}. They are decision-making agents envisioned to learn automatically to take appropriate tactical actions in a timely response to adversarial activities in cyber systems, such as enterprise networks and networked industry systems \cite{John}. Given the complexity of cyber networks, multiple cooperative defender agents, referred to as blue agents, are often considered. As shown in the previous work \cite{survey}, cooperative agents trained using Multi-Agent Reinforcement Learning (MARL) may improve cyber defence effectiveness by coordinating their actions and collectively responding to threats in multi-stage cyber defence operations. However, in a cyber network environment, a large amount of information might be needed to ensure effective decision-making. In fact, the amount of information requested by the human blue team to perform detection analysis and countermeasures is often strictly limited by the network operation centre to prevent it from overwhelming the network. Therefore, in training MARL ACD agents, reducing the needed information in the agents’ observation space is crucial to facilitate training, improve training speed, and reduce information cost during execution. In this work, we apply inter-agent communications to avoid a single large observation space across all agents. This allows the problem space to be decomposed into smaller areas of responsibility for each of the multiple ACD agents. Furthermore, inter-agent communication in MARL enables agents to share information during both training and execution, leading to more robust and coordinated defense strategies \cite{Zhu}. This is akin to what would human experts during incident responses do when they try to avert cyber threats.
This work makes the following contributions:
\begin{itemize}
	\item A novel application of the MARL algorithm with inter-agent communication capabilities for ACD using the Differentiable Inter-Agent Learning (DIAL) algorithm \cite{Forester} adapted for cyber operations.
	\item The first MARL ACD agents to use minimal single-bit inter-agent communication messages that outperform agents that require global state information.
	\item Behavior and performance evaluation of the communicating agents in increasingly complex scenarios, and demonstration of the practical applicability of this approach within an enterprise network simulated in a realistic environment, Cyber Operations Research Gym (CybORG).
\end{itemize}

\Section{MARL Game Design for Cyber Agents}

To generate ACD agents using MARL, the first step is to transform the networked cyber operation use cases into MARL training game episodes using MARL concepts and agent interfaces. The training game consists of the cyber network scenario, agents, their action spaces, observation spaces, and the reward function. In the training games, blue agents (defenders) counter the advances of the red agent (attacker), learning to make defence tactic action decisions and to communicate effectively. Through the MARL training algorithm, blue agents develop their defense strategies, constructed in their decision policy models.

\begin{figure}[ht!]
  \centering
  \includegraphics[width=\linewidth]{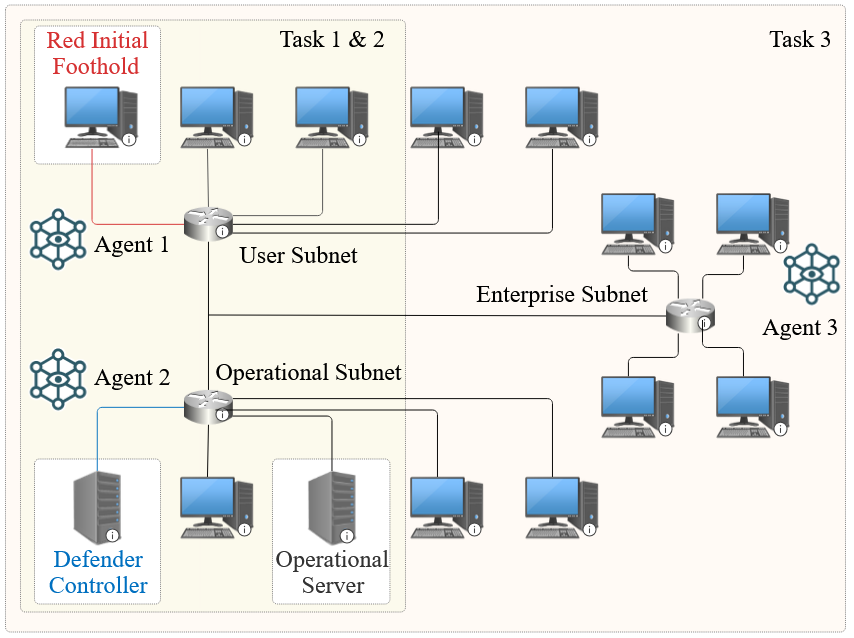}
  \caption{Network configurations in the designed cyber operation games.}
  \label{fig:base_scenario}
\end{figure}

\textbf{Network Configuration}: The games used in this work run on three scenarios: a small network, a small network with a green agent, and a large network. In each training game, its network is simulated in the agent training environment. Figure \ref{fig:base_scenario} depicts the basic network architecture underlying all three scenarios, including the User, Enterprise and Operational subnets.

\textbf{Red Agent}: The goal of the scripted red agent is to infiltrate the server residing in the Operational subnet and disrupt the essential services it provides to all users over the network. The red agent may move in and across the User subnet. After discovering the Enterprise and Operational subnets, the red agent follows the same strategy of actions and attempts to gain admin privileges on hosts. Additionally, should the defender agents succeed with their mitigation actions to remove the red processes, the red agent will attempt to reestablish its sessions on the discovered hosts. Lastly, the red agent can perform the ’impact’ action to stop the key service provided by the Operational server

\textbf{Blue Agents}: One blue agent resides in each subnet to defend against the red attacks. The objective of the blue agents is to prevent the red agent from advancing to the Operational server. 

Cyber operation actions performed by blue agents to protect the network are environmental actions, categorized into passive and active ones. The passive ’monitor’ action, executed at the end of each turn, functions like a host-based IDS, which can generate alerts in response to network connections and process creations, and signal potential malicious activities. The active actions, which must be learned by the agents, include ’remove,’ ’restore,’ ’analyze,’ and ’block,’ each designed to counter specific threats detected within the network.

The reward function considers the importance of different hosts in the network. A host captured by the red agent by installing its privileged shell leads to various penalties: a minor penalty of -0.1 for a User subnet host, a medium penalty of -1.0 for an Enterprise and Operational subnet host, a severe penalty of -10.0 for the Operational server due to the critical service it provides. The ’restore’ action also incurs penalties based on the host’s importance to improve the decision strategy to be learned. Inappropriate actions, such as unnecessary ’remove’ or ’analyze’ actions, incur a standard penalty of -0.5, while a flat penalty of -1.0 is imposed for each ’block’ due to network traffic capacity disruption. The reward function helps defenders prioritize network assets, minimize damage from red agent activities, and adapt their strategies to protect high-value assets.

The observation space vector comprises host and network elements. In the host elements, for each host, the observation vector includes four binary digits: the first two indicate recent red agent activities, such as port scans or exploits, and the last two represent the current host status, ranging from no threats visible to the presence of a privileged shell. The network element of the observation space vector indicates the traffic block condition in the subnet of the blue agent. In the small network scenario, a single block bit indicates whether the traffic to the other subnet is blocked. In the large network scenario, two block bits signify which specific subnet is blocked.

\Section{The Training Approach}

Though both red and blue agent training can be supported in CybORG, this work investigates the training of cooperative strategies for blue agents. The network scenarios, the blue agent’s action and observation spaces and the reward function are all enabled in CybORG using CybORG existing features, or adapting CybORG modules according to the game design.

\textbf{Adapted DIAL Algorithm for Cyber Agents}:
\label{sec: DIAL algo}
In DIAL, during centralized learning, each agent trains its neural network called 'C-Net'. The 'C-Net' outputs both communication and environmental actions. The environmental action is executed, while the communication action, which defines the message to be sent, is inputted to the 'C-Nets' of other agents. Additional inputs to 'C-Net' include the agent’s local observation space, the feedback from other agents on previous messages received from this agent, and the reward resulting from the previous action. 

'C-Net' is a Recurrent Neural Network (RNN) with two hidden layers $h$ that are interconnected and maintained during an episode. For the 'C-Net' architecture, readers are referred to \cite{Forester} for details. Denote the four inputs using a tuple ($o_t^a$, $m_{t-1}^{a'}$, $u_{t-1}^a$, $a$), where $a$ is the agent identifier, $o_t^a$ is the partial observation space of agent $a$ at time $t$, $m_{t-1}^{a'}$ is the message received from agent $a'$ from the previous timestep, and $u_{t-1}^a$ is the previous action of agent $a$. The four inputs are embedded as follows. The embedding of the partial observation space $TaskMLP(o_t^a)$ is generated by a task-specific network which needs to be specific for the environment. For instance, in the Switch Riddle problem \cite{Forester} for which DIAL was designed, $o_t^a$ is passed through a lookup table to produce the embedding $TaskMLP(o_t^a)$. The message input $m_{t-1}^{a'}$ is passed through a 1-layer MLP, and inputs $u_{t-1}^a$ and $a$ are passed through lookup tables to generate their corresponding embeddings. The four components are then summed element-wise to produce the embedding of $z_t^a$ as shown in the following equation. All embeddings have the same size of 128 \cite{Forester}.

\begin{multline}
z_t^a = (TaskMLP(o_t^a) + MLP[|M|, 128](m_{t-1}) + \\
Lookup(u_{t-a}^a) + Lookup(a))
\end{multline}

For our cyber agents defined in the previous section, the above embedding for $m_{t-1}^{a'}$, $u_{t-1}^a$, $a$ can be directly applied. However, since the observation space of each blue agent is much more complex than the previous simple environments, a systematic technique is employed. We first embed the information associated with each host node in the blue agent's observation space by passing through a lookup table. Then, these components are summed element-wise to obtain the final embedding of the agent's observation space. As illustrated in the following equation, $host\_i_t^a, i=1,2,...N$ is the host information and the $blocks_t^a$ is the block bits information in the observation space of agent $a$.

\begin{multline}
z_t^a = ((Lookup(host\_1_t^a) + Lookup(host\_2_t^a) + \dots + \\
Lookup(host\_N_t^a) + Lookup(blocks_t^a)) + \\
MLP[|M|, 128](m_{t-1}) + Lookup(u_{t-a}^a) + Lookup(a))
\end{multline}

$z_t^a$ is then processed through the 2-layer RNN built using Rectified Linear Units (ReLU) and Gated Recurrent Units (GRU), which has similar performance to Long Short-Term Memory (LSTM) \cite{Forester}. The output at timestep $t$ consists of $q$ values for the environment actions to feed to the action selector, and $q$ values for the communication message to feed to a Discretize/Regularize Unit (DRU). 

Agent training begins with randomized communication and action policies. Training then optimizes the policies through iterative trial and error. During training, agents participate in game episodes, interacting with the environment and each other. They optimize decision policies (C-Nets) for actions to counter attackers and for communications with other agents.

\textbf{Mapping C-NET Inter-Agent Communications to CybORG}:
To train cyber agents, in addition to the new embedding scheme for the agent's observation space, the agent communication substrate in the C-Net of the DIAL algorithm must be translated into real message exchanges between agents across the cyber network simulated in CybORG. This is a crucial step in enabling the MARL algorithm, which embeds communication learning within action learning, such as DIAL, in the cyber network training environment. 

At each action step for each agent, the DIAL algorithm generates a "message" as an output from the agent's C-Net in the format of a numeric value in $m \in \mathbb{R}$. During training, $m$ is sent to other agents and used as input to their C-Nets for the next action step. $m$ is received by the C-Net as an input. The C-Net trains to update the agent's policies for both defence actions and communication messages. During execution, each $m$ is discretized to a binary communication message $M$ sent to the agents over the cyber network. When $m=0$, no message needs to be sent. The fewer the message bits required for a better decision model, the more efficient the communications and overall performance of the agents. 


\textbf{Training Agents with Strategic Action Unmasking}: Unless an action is unmasked, it is unavailable in the blue agent's action space, consistent with real-world conditions. For example, an action is unavailable if its execution command cannot be formed due to missing parameters, such as removing a malicious process without knowing the process identifier. At the same time, action masking should not be overused, as it may prevent the agent from discovering its emergent policies.

Therefore, the 'remove' and 'restore' actions are only unmasked upon threat detection. The 'remove' action cannot execute without a threat to eliminate, and the highly disruptive 'restore' action is not permitted in operational networks unless a persistent threat exists on the host or server. The "block" action is always unmasked.

Furthermore, exploration space linking agents' communications and agents' actions is shaped by jointly considering agent communications and action unmasking, referred to as "Strategic Action Unmasking (SAU) ". As per this strategy, the 'analyse' action can only be used by human blue team members upon detecting or receiving an alert of a threat because it is very resource-intensive. Aligning with this real-world condition, the 'analyse' action is contingent upon the detection of a threat shown in the blue agent's observation space. However, since the agent can only see its subnet in the observation space but can communicate with agents in different subnets, the 'analyse' action is also unmasked upon receiving a communication message from other agents. As when, who and what to message is unknown and must be learned by the agents, the condition is based solely on message reception. SAU allows agents to explore integrated decision space on actions and communication messages. Other masked actions are specific to a single host and server, and their masks are not related to inter-agent communications. Therefore, SAU is only applied to the 'analyse' action. 

\Section{Evaluation and results}
Training games run for 30 timesteps per game episode for the small and medium network scenarios and 60 timesteps per episode for the larger network configuration. Each training session consists of 5,000 epochs, with each epoch comprising 128 episodes, resulting in approximately 2 to 4 million training timesteps to construct the agent policies. During evaluation, where the agents execute their learned policies to defend the network, 128 independent episodes are run to calculate the performance metrics. The algorithm parameters are set according to Table \ref{table:dial_hyper}.

\begin{table}[ht!]
\centering
\caption{Parameters for DIAL \& QMix algorithms in CybORG.}
\begin{tabular}{ p{0.35\linewidth} p{0.25\linewidth} p{0.25\linewidth}} 
 \hline
 \textbf{Parameters} & \textbf{DIAL Values} & \textbf{QMix Values}\\ 
 \hline
 Batch size & 128 & 128\\
 Rollout size & 8 & 8 \\
 Learning rate ($\alpha$) & 0.0005 & 0.001 \\ 
 Discount Factor ($\gamma$) & 0.90 & 0.90\\
 Exploration rate start, finish ($\epsilon$) & 1.0, 0.05 & 1.0, 0.05 \\
 Exploration anneal time & 1M timesteps & 1M timesteps\\
 RNN hidden layer dim & 128 & 64 \\
 Target update interval & 100 Epochs & 200 Epochs\\
 \hline
\end{tabular}

\label{table:dial_hyper}
\vspace{-7mm}
\end{table}

\begin{figure}[htb!]
    \subfigure[95\% Detection rate without SAU.]{

        \includegraphics[width=0.75\linewidth, height=5.25cm]{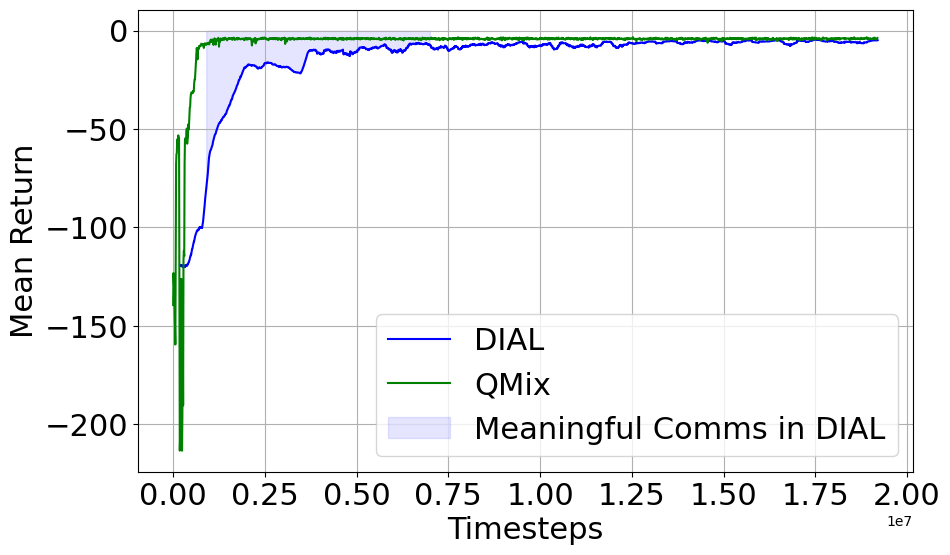} 
}
    \subfigure[50\% Detection rate without SAU.]{

        \includegraphics[width=0.75\linewidth, height=5.25cm]{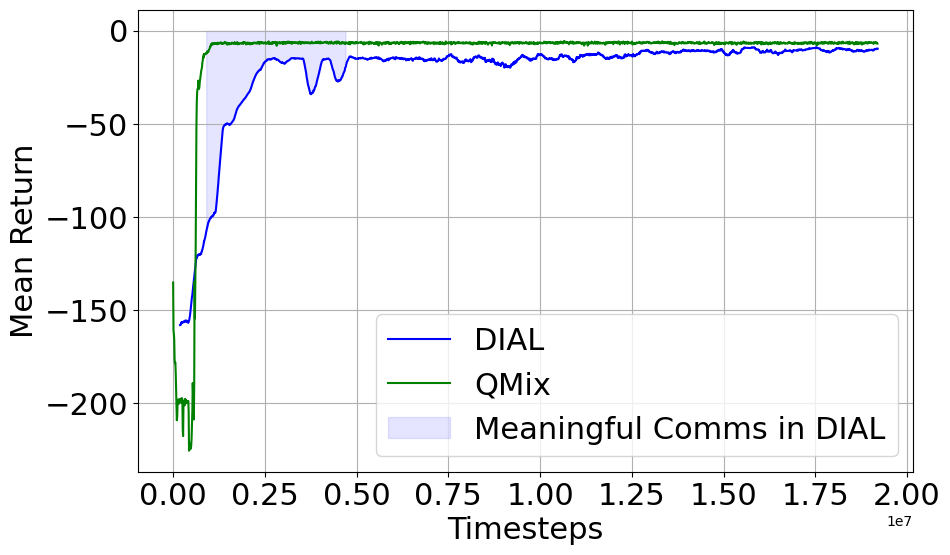} 
}
    \subfigure[50\% Detection rate with SAU.]{

        \includegraphics[width=0.75\linewidth, height=5.25cm]{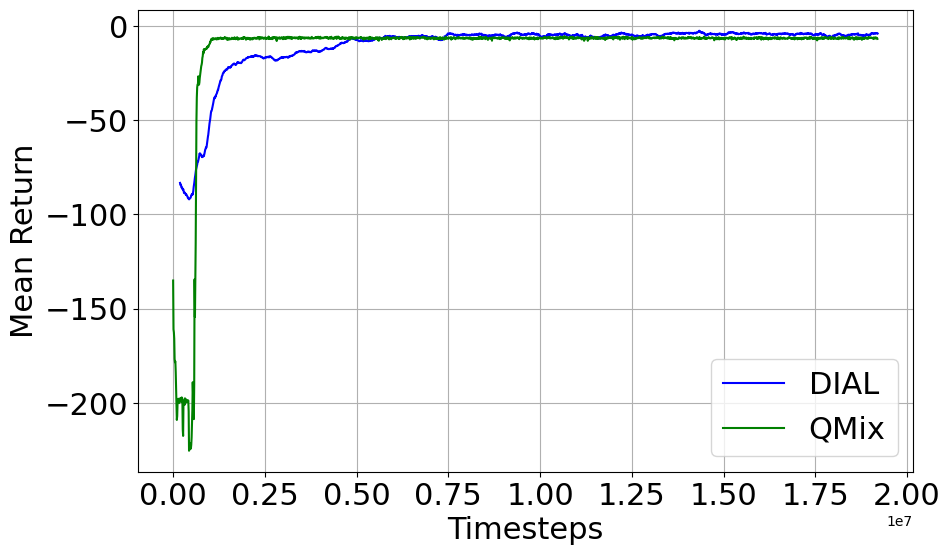} 
}
    \caption{Learning curves from simple game experiments- Small network without 'block' action.}
    \label{fig:learning_curves_small}
\end{figure}

As illustrated in Figure \ref{fig:learning_curves_small}, DIAL employs a distributed architecture where each agent trains using its observation space confined by its subnet, rather than the global state information. While this reduces the transmission cost of collecting global states, it takes longer to learn action decisions and communication messages. This is expected, as DIAL learns additional Q-values for selecting communication messages while learning those for selecting actions. For example, DIAL agents learn more slowly about the usefulness of the 'analyse' followed by the 'restore' action. When the 'monitor' action fails more frequently in detecting the red agent's threats, DIAL is at a greater disadvantage compared to QMix, as shown in \ref{fig:learning_curves_small} (b) versus \ref{fig:learning_curves_small} (a), since QMix learns the general pattern from the global state information. The learning speed and the learned policies of DIAL agents can be improved with SAU which makes 'analyse' available in the agent action space upon receiving a communication message, as revealed in \ref{fig:learning_curves_small} (c). Employing SAU, DIAL agents' policies slightly outperform those of QMix agents, showing the effectiveness of SAU. The mean and
the standard deviation of the return across these experiments are presented in Table \ref{table:eval_scores}.  

During the learning, as shown in Figure \ref{fig:learning_curves_small}, some meaningful communications are observed as DIAL agents did learn to use the 1-bit messaging system to alert potential threats such as port scans from the other subnet. This communication helps agents learn policies initially. However in this setting, the agents have not developed a consistent communication strategy. Increasing the number of message bits does not improve results. DIAL agents show no clear performance advantages in these simple game sets, except for eliminating the high transmission cost of collecting global state information. 

\begin{figure}[htb!]
    \subfigure[Small network.]{

        \includegraphics[width=0.75\linewidth, height=5.25cm]{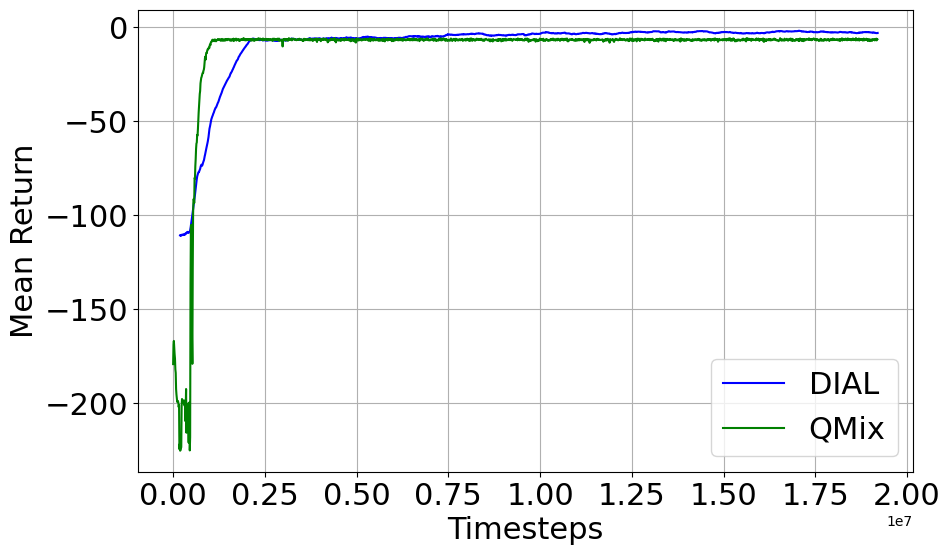} 
}
    \subfigure[Small network with green agent.]{

        \includegraphics[width=0.75\linewidth, height=5.25cm]{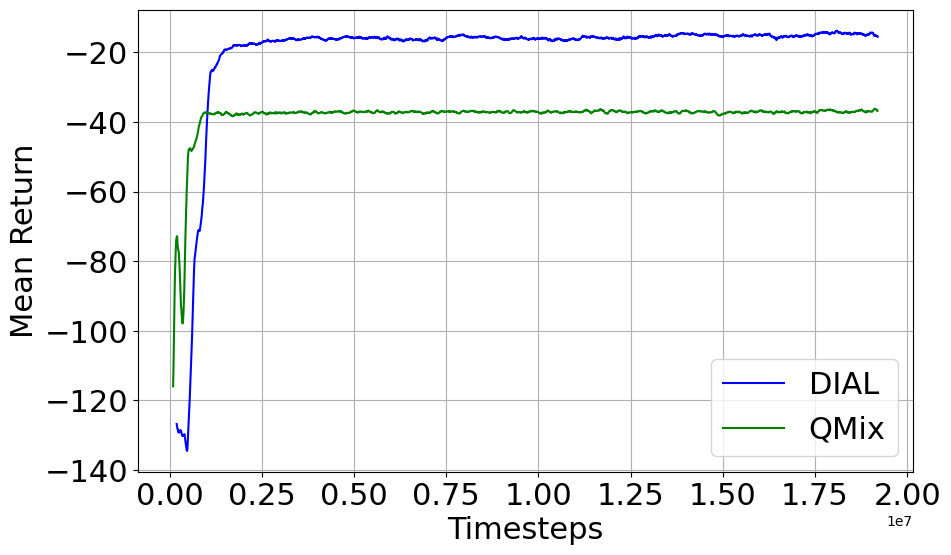} 
}
    \subfigure[Large network.]{

        \includegraphics[width=0.75\linewidth, height=5.25cm]{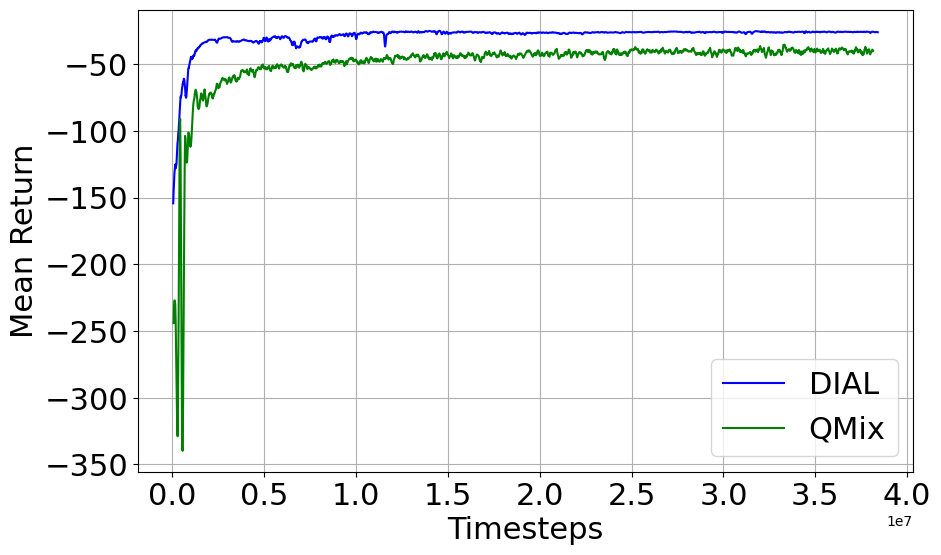} 
}
    \caption{Learning curves for DIAL and QMix in extended games - With SAU and 'block' action.}
    \label{fig:learning_curves}
\end{figure}

In the extended games, the 'block' action and SAU are enabled in all network configurations. The detection rate is set to 50\% to 
 with real-world conditions. The green agent is activated in the small network. As illustrated in Figure \ref{fig:learning_curves}, DIAL agents improve their learning performance across these experiments. The policies learned by DIAL agents also outperform those of QMix agents, as shown in Figure \ref{fig:learning_curves} and Table \ref{table:eval_scores}. DIAL agents utilize a 1-bit message system. Increasing the number of message bits does not improve the performance. This result demonstrates a significant reduction in information transmission costs, indicating that most global state information is unnecessary when agents learn optimal inter-agent communications. 

\begin{table*}[t]
\centering
\caption{Mean returns with standard deviation across experiments.}
\begin{tabular}{c|ccc}
 \hline
 \multicolumn{4}{c}{Simple Game Tests as in Figure \ref{fig:learning_curves_small}} \\
 \hline
 Algorithm & 95\% detection & 50\% detection & 50\% detection\& SAU \\ 
 \hline
 DIAL & $-4.9\pm1.4$ & $-8.2\pm1.5$ & $\boldsymbol{-6.4\pm0.9}$ \\
 QMIX & $\boldsymbol{-4.2 \pm 1.1}$ & $\boldsymbol{-7.3\pm1.2}$ & $-7.1\pm1.3$ \\
 \hline
 \multicolumn{4}{c}{Extended Game Tests as in Figure \ref{fig:learning_curves}: 50\% detection with SAU and `Block' Action} \\
 \hline
 Algorithm & Small network & Small network with green agent & Large network \\ 
 \hline
 DIAL & $\boldsymbol{-3.6\pm0.8}$ & $\boldsymbol{-18.4\pm1.8}$ & $\boldsymbol{-26.4\pm1.5}$\\
 QMIX & $-7.8\pm1.2$ & $-32.7\pm2.3$ & $-43.4\pm4.6$\\
 \hline
\end{tabular}
\label{table:eval_scores}
\end{table*}

DIAL outperforms QMix in these cases by learning to effectively utilize communication messages to coordinate actions like the 'analyse', 'block' and 'restore' when threats are undetected. In small networks, the performance advantage of DIAL over QMix is marginal. However, in larger networks, DIAL demonstrates significant improvement over QMix. Additionally, in small networks, considering a green agent causing false-positive threat detections — a common occurrence in real-world networks — reveals DIAL's superior decision-making compared to QMix. 

The performance advantage of DIAL agents in the presence of green agents enhances the applicability of the ACD agent approach. Green agents simulate general network users, offering valuable insights into the complexities of real-world cyber defence. Network user traffic can mask malicious activities, posing a major challenge for network analysts, who must differentiate between malicious and benign activities to effectively defend the network.

The impact of green agents on blue agents' decision-making is evident from the accumulation of penalties, as shown in Figure \ref{fig:learning_curves} (b). These penalties, while not indicative of actual host compromises, represent the time and resource costs associated with addressing false positives. Such costs are a realistic aspect of cyber operations where not all security alerts correspond to genuine threats but still require investigation and consume valuable resources. Therefore DIAL agents' ability to handle the green agent is crucial.

\section{Conclusion}
This work presented a game design and agent training approach using the MARL system to develop cooperative ACD agents. We have specifically studied the training of agents that can learn optimal policies for cyber defense tactics by leveraging inter-agent communications while simultaneously learning the optimal communication behaviors required to achieve these policies.

\end{document}